\documentclass[tightenlines,superscriptaddress,10pt,twocolumn,aps,prd]{revtex4-1}

\usepackage{geometry}
\usepackage{graphicx,wrapfig}
\usepackage[export]{adjustbox}
\usepackage{epsfig}
\usepackage{subfigure}
\usepackage{sidecap}
\usepackage{notes2bib}
\usepackage{gensymb}
\usepackage{url}
\usepackage{amsmath}
\usepackage{latexsym}
\usepackage{amssymb}
\usepackage{xcolor}
\usepackage{color}
\usepackage{slashed}
\usepackage{units}
\usepackage{siunitx}
\usepackage{notoccite}
\usepackage[utf8]{inputenc}
\usepackage[english]{babel}
\usepackage{csquotes}
\usepackage{dcolumn}
\usepackage{bm}
\usepackage{color}
\usepackage{enumitem}
\usepackage{pifont}
\usepackage{textcomp}
\usepackage{tikz}
\usepackage{colortbl}
\usepackage[version=4]{mhchem}
\usepackage{xspace}
\usepackage{placeins}
\usepackage{tabularx}
\usepackage{hyperref}
\usepackage{amssymb}
\usepackage{hyperref}

\usepackage{pgfgantt}
\usepackage[ampersand]{easylist}

\graphicspath{ {figures/} }

\newcommand{\cevns}{CEvNS}
\newcommand{\about}{{\fontfamily{ptm}\selectfont\texttildelow}}

\newcommand{\logL}{\log{\mathcal{L}}}

\begin{document}

\onecolumngrid

\title{Sensitivity of the COHERENT Experiment to Accelerator-Produced Dark Matter}

\widowpenalty5000
\clubpenalty5000
\renewcommand\floatpagefraction{1}
\renewcommand\topfraction{1}
\renewcommand\bottomfraction{1}
\renewcommand\textfraction{0}

\setlength{\belowcaptionskip}{-10pt} % tighten up float spacing

\renewcommand{\thesection}{\arabic{section}}
\renewcommand{\thesubsection}{\thesection.\arabic{subsection}}
\renewcommand{\thesubsubsection}{\thesubsection.\arabic{subsubsection}}

\makeatletter
\renewcommand{\p@subsection}{}
\renewcommand{\p@subsubsection}{}
\makeatother

\newcommand{\itep}{Institute for Theoretical and Experimental Physics named by A.I. Alikhanov of National Research Centre ``Kurchatov Institute'', Moscow, 117218, Russian Federation}
\newcommand{\mephi}{National Research Nuclear University MEPhI (Moscow Engineering Physics Institute), Moscow, 115409, Russian Federation}
\newcommand{\indiana}{Department of Physics, Indiana University, Bloomington, IN, 47405, USA}
\newcommand{\duke}{Department of Physics, Duke University, Durham, NC 27708, USA}
\newcommand{\tunl}{Triangle Universities Nuclear Laboratory, Durham, NC 27708, USA}
\newcommand{\utk}{Department of Physics and Astronomy, University of Tennessee, Knoxville, TN 37996, USA}
\newcommand{\ornl}{Oak Ridge National Laboratory, Oak Ridge, TN 37831, USA}
\newcommand{\sandia}{Sandia National Laboratories, Livermore, CA 94550, USA}
\newcommand{\fermi}{Enrico Fermi Institute and Kavli Institute for Cosmological Physics, University of Chicago, Chicago, IL 60637, USA}
\newcommand{\chicago}{Department of Physics, University of Chicago, Chicago, IL 60637, USA}
\newcommand{\nmsu}{Department of Physics, New Mexico State University, Las Cruces, NM 88003, USA}
\newcommand{\lanl}{Los Alamos National Laboratory, Los Alamos, NM, USA, 87545, USA}
\newcommand{\cenpa}{Center for Experimental Nuclear Physics and Astrophysics \& Department of Physics, University of Washington, Seattle, WA 98195, USA}
\newcommand{\ncsu}{Department of Physics, North Carolina State University, Raleigh, NC 27695, USA}
\newcommand{\usd}{Physics Department, University of South Dakota, Vermillion, SD 57069, USA}
\newcommand{\virgtech}{Center for Neutrino Physics, Virginia Tech, Blacksburg, VA 24061, USA}
\newcommand{\nccu}{Department of Mathematics and Physics, North Carolina Central University, Durham, NC 27707, USA}
\newcommand{\cmu}{Department of Physics, Carnegie Mellon University, Pittsburgh, PA 15213, USA}
\newcommand{\florida}{Department of Physics, University of Florida, Gainesville, FL 32611, USA}
\newcommand{\laurentian}{Department of Physics, Laurentian University, Sudbury, Ontario P3E 2C6, Canada}
\newcommand{\tufts}{Department of Physics and Astronomy, Tufts University, Medford, MA 02155, USA}
\newcommand{\kaist}{Department of Physics at Korea Advanced Institute of Science and Technology (KAIST)
and Center for Axion and Precision Physics Research (CAPP) at Institute for Basic Science (IBS), Daejeon, 34141, Republic of Korea}

\author{D.~Akimov}\affiliation{\itep}\affiliation{\mephi}
%\author{J.B.~Albert}\affiliation{\indiana}
\author{P.~An}\affiliation{\duke}\affiliation{\tunl}
\author{C.~Awe}\affiliation{\duke}\affiliation{\tunl}
\author{P.S.~Barbeau}\affiliation{\duke}\affiliation{\tunl}
\author{B.~Becker}\affiliation{\utk}
\author{V.~Belov}\affiliation{\itep}\affiliation{\mephi}
\author{M.A.~Blackston}\affiliation{\ornl}
\author{A.~Bolozdynya}\affiliation{\mephi}
\author{B.~Cabrera-Palmer}\affiliation{\sandia}
%\author{M.~Cervantes}\affiliation{\duke}
\author{N.~Chen}\affiliation{\cenpa}
\author{E.~Conley}\affiliation{\duke}
\author{R.L.~Cooper}\affiliation{\nmsu}\affiliation{\lanl}
\author{J.~Daughhetee}\affiliation{\utk}
\author{M.~del~Valle~Coello}\affiliation{\indiana}
\author{J.A.~Detwiler}\affiliation{\cenpa}
%\author{M.~D'Onofrio}\affiliation{\indiana}
\author{M.R.~Durand}\affiliation{\cenpa}
\author{Y.~Efremenko}\affiliation{\utk}\affiliation{\ornl}
%\author{E.M.~Erkela}\affiliation{\cenpa}
\author{S.R.~Elliott}\affiliation{\lanl}
\author{L.~Fabris}\affiliation{\ornl}
\author{M.~Febbraro}\affiliation{\ornl}
\author{W.~Fox}\affiliation{\indiana}
\author{A.~Galindo-Uribarri}\affiliation{\utk}\affiliation{\ornl}
\author{M.P.~Green}\affiliation{\tunl}\affiliation{\ornl}\affiliation{\ncsu}
\author{K.S.~Hansen}\affiliation{\cenpa}
\author{M.R.~Heath}\affiliation{\ornl}\affiliation{\indiana}
\author{S.~Hedges}\affiliation{\duke}\affiliation{\tunl}
\author{T.~Johnson}\affiliation{\duke}\affiliation{\tunl}
\author{M.~Kaemingk}\affiliation{\nmsu}
\author{L.J.~Kaufman}\altaffiliation[Now at: ] {SLAC National Accelerator Laboratory, Menlo Park, CA 94205, USA}\affiliation{\indiana}
\author{A.~Khromov}\affiliation{\mephi}
\author{A.~Konovalov}\affiliation{\itep}\affiliation{\mephi}
\author{E.~Kozlova}\affiliation{\itep}\affiliation{\mephi}
\author{A.~Kumpan}\affiliation{\mephi}
\author{L.~Li}\affiliation{\duke}\affiliation{\tunl}
\author{J.T.~Librande}\affiliation{\cenpa}
\author{J.M.~Link}\affiliation{\virgtech}
\author{J.~Liu}\affiliation{\usd}
\author{K.~Mann}\affiliation{\tunl}\affiliation{\ornl}
\author{D.M.~Markoff}\affiliation{\tunl}\affiliation{\nccu}
\author{H.~Moreno}\affiliation{\nmsu}
\author{P.E.~Mueller}\affiliation{\ornl}
\author{J.~Newby}\affiliation{\ornl}
\author{D.S.~Parno}\affiliation{\cmu}
\author{S.~Penttila}\affiliation{\ornl}
\author{D.~Pershey}\affiliation{\duke}
\author{D.~Radford}\affiliation{\ornl}
\author{R.~Rapp}\affiliation{\cmu}
\author{H.~Ray}\affiliation{\florida}
\author{J.~Raybern}\affiliation{\duke}
\author{O.~Razuvaeva}\affiliation{\itep}\affiliation{\mephi}
\author{D.~Reyna}\affiliation{\sandia}
\author{G.C.~Rich}\affiliation{\fermi}
\author{D.~Rudik}\affiliation{\itep}\affiliation{\mephi}
\author{J.~Runge}\affiliation{\duke}\affiliation{\tunl}
\author{D.J.~Salvat}\affiliation{\indiana}
\author{K.~Scholberg}\affiliation{\duke}
\author{A.~Shakirov}\affiliation{\mephi}
\author{G.~Simakov}\affiliation{\itep}\affiliation{\mephi}
\author{G.~Sinev}\affiliation{\duke}
\author{W.M.~Snow}\affiliation{\indiana}
\author{V.~Sosnovtsev}\affiliation{\mephi}
\author{B.~Suh}\affiliation{\indiana}
\author{R.~Tayloe}\affiliation{\indiana}
\author{K.~Tellez-Giron-Flores}\affiliation{\virgtech}
\author{R.T.~Thornton}\affiliation{\indiana}\affiliation{\lanl}
\author{I.~Tolstukhin}\altaffiliation[Now at: ] {Argonne National Laboratory, Argonne, IL 60439, USA}\affiliation{\indiana}
\author{J.~Vanderwerp}\affiliation{\indiana}
\author{R.L.~Varner}\affiliation{\ornl}
\author{C.J.~Virtue}\affiliation{\laurentian}
\author{G.~Visser}\affiliation{\indiana}
\author{C.~Wiseman}\affiliation{\cenpa}
\author{T.~Wongjirad}\affiliation{\tufts}
\author{J.~Yang}\affiliation{\tufts}
\author{Y.-R.~Yen}\affiliation{\cmu}
\author{J.~Yoo}\affiliation{\kaist}
\author{C.-H.~Yu}\affiliation{\ornl}
\author{J.~Zettlemoyer}\affiliation{\indiana}

\begin{abstract}
The COHERENT experiment is well poised to test sub-GeV dark matter models using low-energy recoil detectors sensitive to coherent elastic neutrino-nucleus scattering (\cevns) in the $\pi$-DAR neutrino beam produced by the Spallation Neutron Source.  We show how a planned 750-kg liquid argon scintillation detector would place leading limits on scalar light dark matter models, over two orders of magnitude of dark matter mass, for dark matter particles produced through vector and leptophobic portals in the absence of other effects beyond the standard model.  The characteristic timing structure of a $\pi$-DAR beam allows a unique opportunity for constraining systematic uncertainties on the standard model background in a time window where signal is not expected, enhancing expected sensitivity.  Additionally, we discuss future prospects, further increasing the discovery potential of \cevns\ detectors.  Such methods would test the calculated thermal dark matter abundance for all couplings $\alpha'\leq1$ within the vector portal model over an order of magnitude of dark matter masses.
\end{abstract}

\twocolumngrid

\maketitle

\pagenumbering{arabic}
\pagestyle{plain}

\section{Introduction}
\label{sec:Theory}

\begin{figure*}[!bt]
\centering
\includegraphics[width=\linewidth]{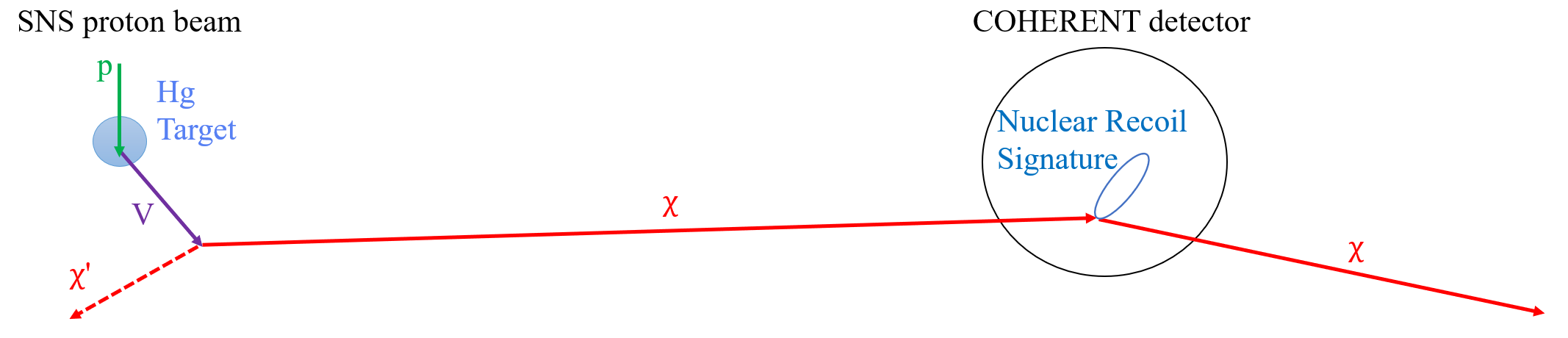}
\caption{Overview of a postulated accelerator-produced DM scatter produced at the Spallation Neutron Source (SNS) beam and measured by COHERENT.  A beam of protons is incident on a mercury target on the left.  In each p-Hg interaction, a portal particle, $V$, to the dark sector may be produced.  This portal particle would then decay into a pair of dark matter particles, $\chi\chi'$.  Each of these may then interact in a nearby COHERENT detector at a large off-axis angle to the beam.}
\label{fig:cartoon}
\end{figure*}

Evidence from cosmology continues to strengthen the conclusion that \about\SI{80}{\percent} of the matter in the universe is not composed of standard model (SM) particles~\cite{Freese:2017idy}.  While the gravitational force of this dark matter is readily observed over a large range of cosmological distance scales, its particle nature remains elusive.

Detectors sensitive to low-energy nuclear recoils are ideal for searching for sub-GeV particles postulated as dark matter candidates~\cite{deNiverville:2015mwa,Ge:2017mcq,Abdullah:2018ykz,Dutta:2019fxn}. Similar to the \cevns\ process~\cite{Akimov:2017ade}, a dark matter particle can interact coherently with an entire target nucleus at a momentum transfer of $Q^2 < (50$ MeV)$^2$, greatly enhancing the scattering cross-section. Thus, a relatively small (tonne or few tens of tonne scale) detector sensitive to the several-keV nuclear recoil typical of \cevns\ can surpass the sensitivity of kilotonne-scale detectors.

To satisfy the Lee-Weinberg bound for the WIMP mass~\cite{PhysRevLett.39.165}, sub-GeV dark matter models must also predict a ``portal'' particle to mediate interactions between the relic dark matter candidate and Standard Model particles.  Such a weakly-coupled dark portal particle could be produced at the SNS by decay of $\pi^0$/$\eta^0$ particles and nuclear absorption of $\pi^-$ particles produced by the interactions between the 1 GeV proton beam and mercury target.  The portal particle would subsequently decay to a pair of scalar dark matter particles, either of which may interact within a detector~\cite{deNiverville:2015mwa}. This is shown schematically in Fig.~\ref{fig:cartoon}.

A detector sensitive to \cevns{}-like processes would constrain light dark matter.  Two models are used to benchmark the COHERENT sensitivity distinguished by the nature of the mediator between SM and dark matter particles: a model with a vector portal that kinetically mixes with the photon~\cite{BOEHM2004219,Fayet_2004,deNiverville:2011it}, and a model with a leptophobic portal~\cite{Batell_2014} coupling to any SM baryon.  In addition to portal and dark matter particle masses, $m_V$ and $m_\chi$, the vector portal model has two coupling constants as free parameters, $\epsilon$ and $\alpha'$,  while the leptophobic parameter depends on a single $\alpha_B$.  The parameters of the vector portal model can be conveniently compared to the cosmological density of dark matter through the dimensionless quantity, $Y$~\cite{Izaguirre:2015yja}, given by
\begin{equation}
Y=\epsilon^2\alpha'\left(\frac{m_\chi}{m_V}\right)^4.
\label{eq:dmscaling}
\end{equation}
For a chosen $m_V$ and $m_\chi$, The event rate expected for coherent $\chi$-nucleus scattering scales like $Y^2$ while the cosmological density scales like $Y^2/\alpha'$.  As such, it is conservative to test the values of $\alpha'$ near the perturbative limit.  Current limits fall short of the flux needed to explain cosmologically observed dark matter, and thus more effort is needed to test this model.  A recent analysis of released COHERENT CsI data \cite{Dutta:2019nbn} hints at a roughly $2\sigma$ excess in the region where dark matter scatters would be expected, suggesting this is an exciting area to pursue.

In the leptophobic model, it is more difficult to connect $\alpha_B$ to the expected dark matter relic density, as the relationship is model dependent~\cite{deNiverville:2016rqh}. We thus quote our sensitivity to this model in terms of $\alpha_B$.  Our proposed sensitivity to the leptophobic portal is interesting as beam dump experiments are frequently most sensitive to $\nu-e$ elastic scattering~\cite{PhysRevD.63.112001,deNiverville:2011it}, which is incapable of testing this model.  MiniBooNE was able to constrain this model through specialized beam-stop running which mitigated the high-energy neutrino backgrounds \cite{PhysRevD.98.112004,Aguilar-Arevalo:2017mqx}.

\section{The COHERENT Experiment at the Spallation Neutron Source}
\label{sec:Experiment}
The Spallation Neutron Source (SNS) located at Oak Ridge National Laboratory  (ORNL) is the world's premier facility for neutron-scattering research, producing pulsed neutron beams with intensities an order of magnitude larger than any other currently-operating facility. At full beam power, approximately \num{1.5e14}, \SI{1.01}{\GeV} protons bombard the liquid mercury target in short \SI{600}{\ns}~wide bursts at a rate of \SI{60}{\hertz}. Neutrons produced in spallation reactions with the mercury target thermalize in cryogenic moderators surrounding the target and are delivered to neutron-scattering instruments in the SNS experiment hall. The SNS is a user facility and operates approximately two-thirds of the year.

As a byproduct, the SNS also provides the world's most intense pulsed source of neutrinos in an energy region of specific interest for particle and nuclear astrophysics \cite{AvignoneIII_2003}. Interactions of the proton beam in the mercury target produce $\pi^+$ and $\pi^-$ in addition to neutrons. The pion decay-in-flight fraction is very small, and more than 99$\%$ of $\pi^-$ are absorbed due to the high-$Z$ of the target. This $\pi^+$ decay-at-rest ($\pi$-DAR) neutrino flux therefore has very small uncertainties on spectral shape, timing, and flavor.

The sharp SNS beam timing structure is highly beneficial for background rejection and precise characterization of those backgrounds not associated with the beam~\cite{Bolozdynya:2012xv}.  Looking for beam-related signals only in the \SI{10}{\micro\s} window after a beam spill imposes a factor of $\sim$2000 reduction in the steady-state background, with mitigation for the $\nu_\mu$ flux coincident with the beam pulse better by an order of magnitude.  COHERENT detectors are placed in a basement hallway, called ``Neutrino Alley", where the the neutron flux from the SNS is known to be low, reducing beam-related neutron backgrounds in COHERENT analyses.  We assume 1.4 MW running with $1.5\times10^{23}$ protons-on-target accumulated per year.  With a planned second target station at the SNS, this could increase to 2.4 MW \cite{osti_1185891} providing quicker accumulation of exposure.

COHERENT collaboration deployments of low-energy recoil detectors in Neutrino Alley include~\cite{Akimov:2018ghi}: a CsI[Na] detector (used for the ``first light" \cevns\ measurement~\cite{Akimov:2017ade}, for which data-taking is now complete), a 24 kg liquid argon detector, ``CENNS-10" \cite{Tayloe_2018} ( \cite{Akimov:2019rhz} with low-threshold analysis underway), a 185 kg NaI[Tl] detector, currently operating in high-threshold mode (but with plans to expand to 3.3~tonnes and lower the threshold), and 16 kg of planned HPGe PPC detectors.  All of these will have some sensitivity to accelerator-produced dark matter. Liquid argon scintillation detectors have been built on the several-tonne scale with a relatively low threshold. This scalability makes liquid argon the most promising detector for constraining sub-GeV dark matter and is thus the focus of this paper.  Currently, we plan to deploy a tonne-scale liquid argon detector, LAr-1t, in Neutrino Alley to significantly improve on the success of CENNS-10.  A proposed cryogenic NaI[Ti] scintillation detector in Neutrino Alley would likely improve constraints on dark matter masses below 1~MeV$/c^2$, though a more thorough understanding of the technology is needed. 

The LAr-1t detector is targeted to replace CENNS-10 in Neutrino Alley.  A cylindrical cryostat will house 750 kg of chilled argon (610 kg fiducial).  A coating of TPB will shift the ultra-violet scintillation to wavelengths compatible with the photo-detectors.  The detector will be instrumented with either PMT or SiPM photo-detectors with SiPM detectors offering a lower threshold.  The detector will be shielded with water and lead to reduce the neutron and gamma activity within the cryostat.   Additional shielding could be added in Neutrino Alley to close a hallway that allows open-air access to the beam hall.  This would further reduce the neutron background.

\section{A Dark Matter Search with LAr-1t}
\label{sec:Analysis}
We present an estimate of the LAr-1t sensitivity to constrain the vector and leptophobic portals to scalar dark matter and discuss potential improvements for future detectors using the same detection strategy.  Only dark matter particle masses $1 < m_\chi < 100$ MeV$/c^2$ are considered.  Lower-mass dark matter is currently inaccessible due to the detector threshold, though proposed detectors with a $\sim$keVnr threshold may expand the testable parameter space.  At $Q>\hbar c$~/~1~fm, coherent nuclear recoils are highly suppressed through the nuclear form factor \cite{Payne:2019wvy}.  For portal masses greater than 100~MeV$/c^2$, fewer than 25$\%$ of nuclear scatters have a suitably low $Q^2$ to avoid this suppression.  Liquid argon may still be competitive in this region through inelastic scattering, but more study is needed.  

Signal efficiencies and background rates are estimated from our operating experience with CENNS-10 which together yield a prediction as a function of recoil energy and arrival time relative to the SNS pulse.  We also incorporate systematic errors on this prediction.  Systematic nuisance parameters are profiled by maximizing a log-likelihood prediction while incorporating dark matter.  The fit is energy-dependent to take advantage of differences in the recoil spectrum of \cevns\ and dark matter events.  For this calculation, all beyond the standard model effects, such as neutrino non-standard interactions, outside of the hypothesized dark matter and mediator particles are ignored.  

\subsection{Dark Matter Scatters}

\label{sect:DMSearches}

A flux of dark matter particles passing through LAr-1t would interact coherently with an argon nucleus, producing a rate of nuclear recoil signatures.  We use the BdNMC event generator \cite{deNiverville:2016rqh} to determine the energy spectrum of argon recoils in our detectors, parameterized by the dark matter and portal masses.  Dark matter is generated by decaying portal particles produced through three processes: $\pi^0$ decay, $\eta^0$ decay, and nuclear absorption of $\pi^-$.  The $\pi^0$($\eta^0$) decay channel does not contribute for $m_\chi > m_\pi/2(m_\eta/2)$ due to kinematic constraints.  A GEANT4 \cite{Geant4Cite} simulation of the SNS is used to determine the pion kinematics relevant for determining the dark matter flux.  The simulation predicts 0.09~$\pi^+$/$\pi^0$, 0.04~$\pi^-$, and 0.002~$\eta$ per incident proton are produced at $T_p = 1.01$ GeV \cite{Rapp:2019vnv}, the current SNS beam energy.

Though this is below the $pp\rightarrow pp\eta$ threshold, $T_p = 1.25$ GeV for a stationary proton target, nuclear motion within atoms in the target allows for production at the SNS beam energy.  A calculation that explicitly accounts for sub-threshold production \cite{Cassing1991} agrees with the BdNMC prediction to 30$\%$.

The argon recoil spectrum is then convolved with detector resolution effects.  Applying the argon quenching factor (ratio of detector response for nuclear recoils with respect to electron energy loss) and light yield results in the distribution of observed energy deposited in the detector.  The quenching factor is assumed linear and fit to several independent measurements \cite{aris2018,scene2015,creus2015lar,gastler2012lar}; it is between 25 and 30$\%$ in the region of interest.  This is then smeared using an energy-dependent resolution that is $\sim$18$\%$ at 20 keV$_\text{nr}$ near the center of the \cevns\ distribution.  We assume a light yield of 4.2 photoelectrons per keVee, which gives an efficiency curve with a 20~keVnr threshold.  We have achieved this light yield with the CENNS-10 prototype detector and a detector simulation shows this is achievable with LAr-1t.

For the baseline between the target and detector, $L=28.4$~m, the time of flight for dark matter particles is on average 3 ns longer than $L/c$, much less than the SNS beam pulse width of 600 ns.  Further, no simulated events reached the detector more than 1~$\mu$s after the arrival of the first neutrinos from the beam.  Thus, the dark matter signal is coincident with the ``prompt" $\nu_\mu$ flux produced from $\pi^+$ decay in the SNS target.

\subsection{Analysis Backgrounds}

\begin{figure}[!bt]
\centering
\includegraphics[width=0.49\textwidth]{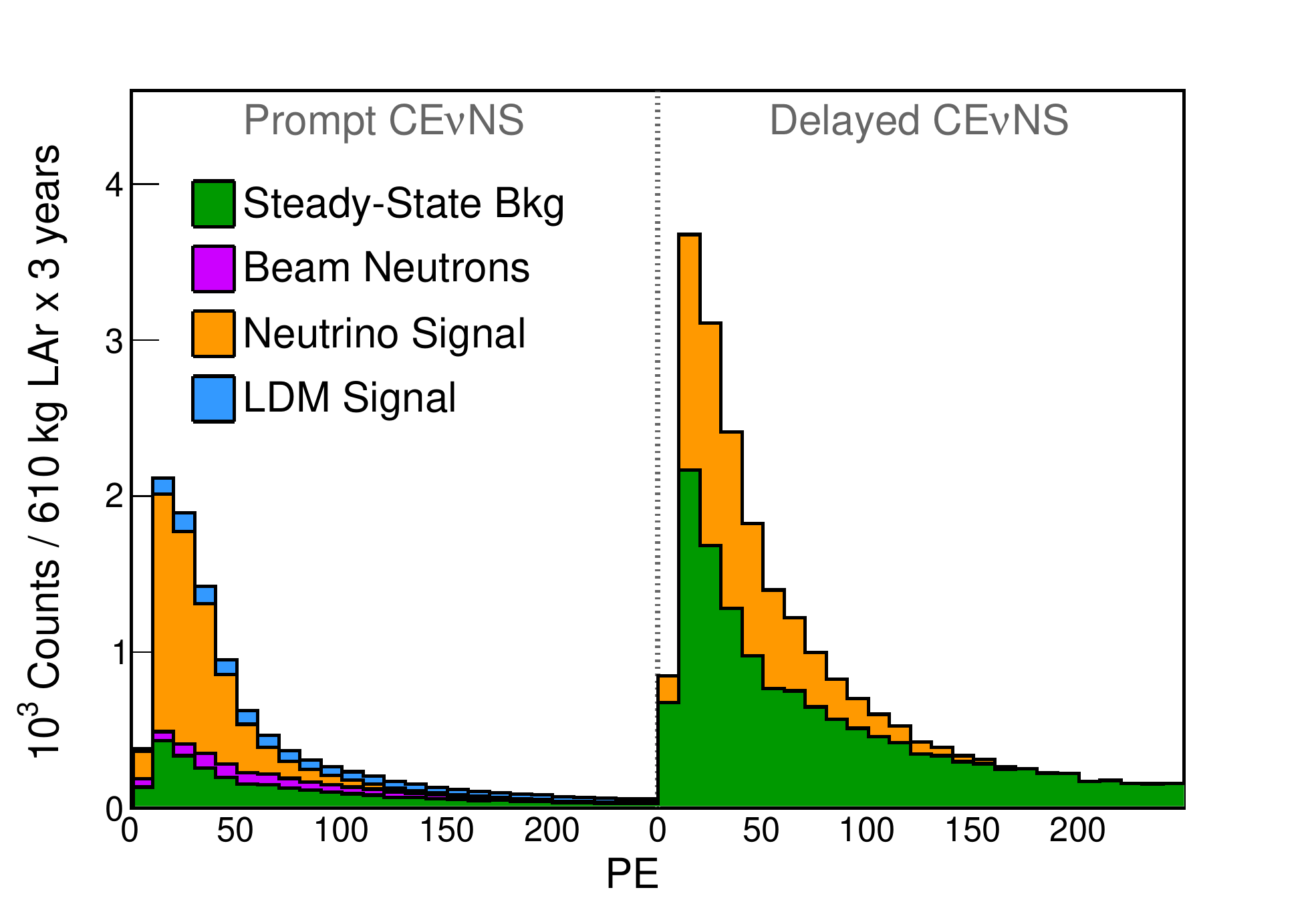}
\caption{The predicted distribution of expected dark matter signal, blue, and analysis backgrounds expected in three years with 610 kg of fiducial volume.  Prompt events are those that appear within the first 1 $\mu$s of the waveform while delayed events reconstruct between 1 and 6 $\mu$s.  The signal lies entirely within the 1 $\mu$s prompt window, allowing for in-situ constraint of \cevns\ uncertainties in the 5$\mu s$ delayed time bin.  The steady-state background will be measured from out-of-time data and assumes filling with underground argon.  The neutron background estimate will depend on additional measurements for a precise prediction.  The dark matter component assumes the vector portal model with $\epsilon=8.77\times10^{-5}$, $\alpha'=0.5$, $m_\chi=15$~MeV/$c^2$, and $m_V=45$~MeV/$c^2$.}
\label{fig:DMPredictionAr}
\end{figure}

There are three sources of background for this search, each of which has already been studied in CENNS-10 \cite{Akimov:2019rhz}: steady-state backgrounds, beam-related neutrons, and \cevns.  The expected energy distribution of backgrounds, along with dark matter signal normalized to $4\times$ the sensitivity limit, is shown in Fig. \ref{fig:DMPredictionAr}.

Steady-state backgrounds are measured in-situ using out-of-time data taken while the beam is on.  Thus, the distribution of these events is known without bias.  Though directly measured, this background reduces our statistical sensitivity due to the high rate, primarily from $\beta$ decay of  $^{39}$Ar.  This is a cosmogenically-activated isotope of argon with a lifetime of 269 years.  The concentration of $^{39}$Ar can be dramatically reduced by filling the detector with argon mined from underground \cite{Galbiati:2007xz}.  For every 1~$\mu$s of livetime integrated per spill, we expect 14.58~(1.52) events for each kg-year of exposure for atmospheric (underground) argon.  For running with underground argon, we expect a 100$\times$ reduction of the $^{39}$Ar background.  We assume no change to the steady-state background related to other sources, which is conservative as roughly 10$\%$ of the background is produced by a pipe of radioactive gas that runs through Neutrino Alley.  We are currently installing a shielding to mitigate this background, though its effectiveness has not been fully determined.  As this background does not introduce a systematic error into the analysis, we can compensate for the increased $^{39}$Ar rate in atmospheric argon by increasing exposure.  Roughly a 3.1$\times$ increase in exposure is needed to match sensitivities calculated in the case of underground argon if LAr-1t is filled with atmospheric argon.

We will also have a beam-related background, dominated by neutrons.  The neutron flux through Neutrino Alley is low enough that a \cevns\ measurement is possible, though neutrons still yield a notable background component for these analyses \cite{Akimov:2017ade,Akimov:2019rhz}.  These scatters are prompt, occurring concurrently with the $\nu_\mu$ flux through the detector.  There is a secondary flux of neutrons stemming from Neutrino-Induced Neutrons (NINs) where a neutron is emitted in a neutrino interaction with a neutron-rich atom \cite{PhysRevD.67.013005} such as lead used in the detector shielding.  The NIN background is largely uncorrelated with the uncertainty in the prompt neutron background and is not confined to the prompt time window.  However, fewer than one event per LAr-1t running in the entire 6 $\mu$s analysis window is expected to penetrate the water shielding surrounding the detector and produce a signal so that the uncertainty in this background has a negligible impact on the analysis.

With uncertainties in both the normalization, energy distribution, and time profile of the neutron background, an ancillary measurement of the energy-dependent neutron flux at the detector location will be essential.  As the majority of neutrons are prompt, and thus coincident with signal, a significant bias in this background rate could result in a false positive detection of dark matter.  The prompt neutron rate is estimated to be 0.53 events for each kg-year of exposure.  This is a significantly reduced rate compared to our experience with CENNS-10.  A GEANT4 simulation suggests this rate is achievable with shielding the detector with a neutron moderator though must be tested in-situ.  The projected sensitivity does not strongly depend on the neutron rate, however.  A neutron normalization 10$\times$ higher would only degrade the projected sensitivity on $Y$ by a factor of roughly 1.8, assuming the neutron event rate is known to the same precision.

Finally, \cevns\ will give a significant background with 2.98 (4.07) selected \cevns\ per kg-year in the prompt (delayed) timing window.  The expression for the \cevns{} cross section is cleanly predicted in the SM, however, there is an uncertainty associated with the argon form factor suppression of the cross section.  The effect of this uncertainty on our predicted \cevns{} rates is much smaller than other error sources.  Additionally, the uncertainties on this background in the prompt window can be constrained with data through analysis of the delayed \cevns .  As there are strong correlations between prompt and delayed events, a simultaneous fit of prompt and delayed events will mitigate the systematic uncertainty on the prompt neutrino background.

\subsection{Systematic Uncertainties}

We have evaluated the errors associated with the likely leading sources of systematic uncertainty for this study, incorporating both normalization and shape differences into the analysis.  We assign a 10$\%$ uncertainty on the SNS flux, estimated by a GEANT4 simulation\cite{Geant4Cite} of the beam, by taking the spread in neutrino yields determined by several physics models.  This same uncertainty is applied to the dark matter flux produced through $\pi^0$ decay and $\pi^-$ absorption.  A 30$\%$ uncertainty on the dark matter flux produced by $\eta^0$ decay is included and uncorrelated to cover differences between the $\eta$ production estimates listed in Sec. \ref{sect:DMSearches}.      

There is also an uncorrelated 10$\%$ uncertainty on the neutron flux.  This is the target precision for forthcoming neutron flux measurements within Neutrino Alley.  Understanding the neutron flux to only 50$\%$ would reduce our sensitivity on $\phi_{DM}$ by up to a factor of two while a reduced uncertainty of 5$\%$ or lower would further improve our constraint.

We consider two additional sources of uncertainty that affect the \cevns{} and dark matter recoil spectra.  A roughly 2$\%$ uncertainty on the quenching factor in liquid argon, motivated by external quenching factor measurements~\cite{aris2018,scene2015,creus2015lar,gastler2012lar}, adjusts the shape of the distribution of ionization energy deposited by \cevns{} and signal events given the distribution of nuclear recoil energies.  Integrated over the energy spectrum, the quenching factor gives a 1.2$\%$ uncertainty on the \cevns{} event count.  We also account for uncertainty in the \cevns{} cross section by adjusting the nuclear form factor.  We assume the Helm form factor formula, and adjust the neutron radius by $\pm3\%$ from the central value, whuch is slightly more conservative than the spread from a recent calculation for argon \cite{Payne:2019wvy}.  This gives a 1.5$\%$ uncertainty in the total event rate.  Both the quenching factor and form factor uncertainties apply to \cevns{} and dark matter recoils, and we treat these effects as correlated for both event types.  There is no systematic uncertainty included for the timing of the neutrino pulse through Neutrino Alley,
under the realistic assumption that pulse arrival time determination will be done with sufficient precision to make this a negligible effect.

As the dark matter signal is relativistic and falls within the prompt time window, there is no signal but several thousand \cevns\ expected in the delayed sample, making the delayed sample a convenient sideband for constraining systematic uncertainties.  The time structure and energy distribution of the neutrino flux is set by the kinematics of $\pi^+$ and $\mu^+$ decay, which are very well understood, and the decay-in-flight contamination to the flux is less than a 1$\%$ contribution.  Thus, uncertainties that affect the \cevns{} background in the prompt time window are strongly correlated with any \cevns{} seen in the delayed window, and data observed in the delayed window can constrain the standard model background expected in the prompt window, coincident with dark matter signal.  

\begin{figure}[!bt]
\centering
\includegraphics[width=0.49\textwidth]{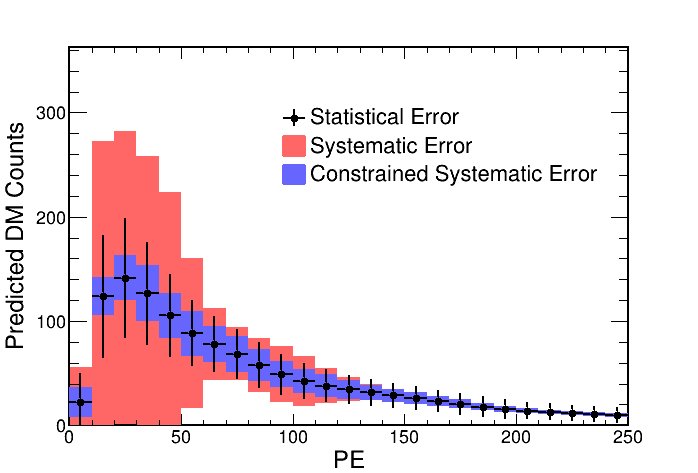}
\caption{The estimated error band on the dark matter prediction for three years of LAr-1t running within the leptophobic model with $m_\chi$ = 5~MeV$/c^2$, $m_V$ = 50~MeV$/c^2$, and $\alpha_B$ set to 50$\%$ higher than the sensitivity limit.  Statistical errors are given by the Poisson error of subtracting the background.  The red error band gives the total systematic error band.  The blue band gives the systematic error after constraining the errors with delayed \cevns\ events.}
\label{fig:DMErrorBand}
\end{figure}

To illustrate the power of this delayed constraint, the recoil spectrum of the background-subtracted dark matter prediction is shown along with the systematic error band in Figure \ref{fig:DMErrorBand}.  Also shown is the systematic error band after constraining the prompt background with events in the delayed bin.  There is a significant reduction of the systematic uncertainty, particularly near the peak of the distribution where the unconstrained systematic error band is larger than the signal prediction.

\subsection{Sensitivity Calculation}

For a given $m_\chi$ and $m_V$, we calculate the minimum dark matter coupling constants that are inconsistent with the Asimov prediction~\cite{Cowan:2010js} without dark matter.  For any choice of model parameters and nuisance parameters, a $-2\logL$ is defined by
\begin{equation}
\begin{split}
    -2\logL(\vec{\theta}) = \sum_{i = 1}^{N_\text{bins}}\left[e_i-o_i+o_i\log{\frac{o_i}{e_i}}\right] + \sum_{j = 1}^{N_\text{syst}}\frac{\delta_j^2}{\sigma_j^2}
\end{split}
\end{equation}
with $\vec{\theta}$ giving the model parameters and $\vec{\delta}$ giving the pulls on each systematic uncertainty.  There are 24~bins in the sum: 12~energy bins in both the prompt and delayed time window.  The expected counts, $e_i$, is a function of $\vec{\theta}$ and $\vec{\delta}$, and $o_i$ gives the Asimov prediction in each bin used as fake data.  A $\chi^2=-2\logL$ is calculated for all possible values of $\vec{\theta}$ by profiling over all nuisance parameters.

\begin{figure}[!bt]
\centering
\includegraphics[width=0.47\textwidth]{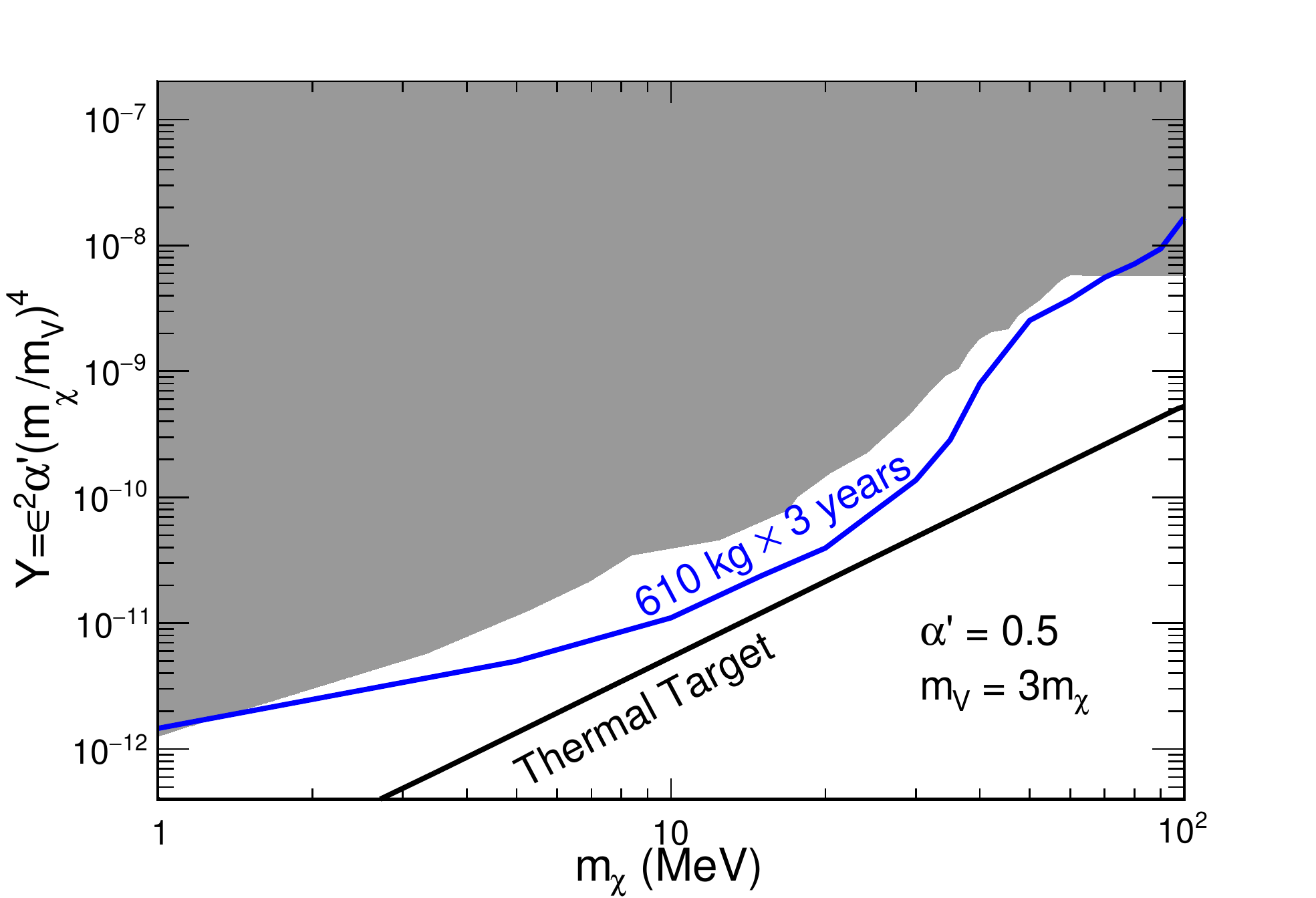}
\includegraphics[width=0.47\textwidth]{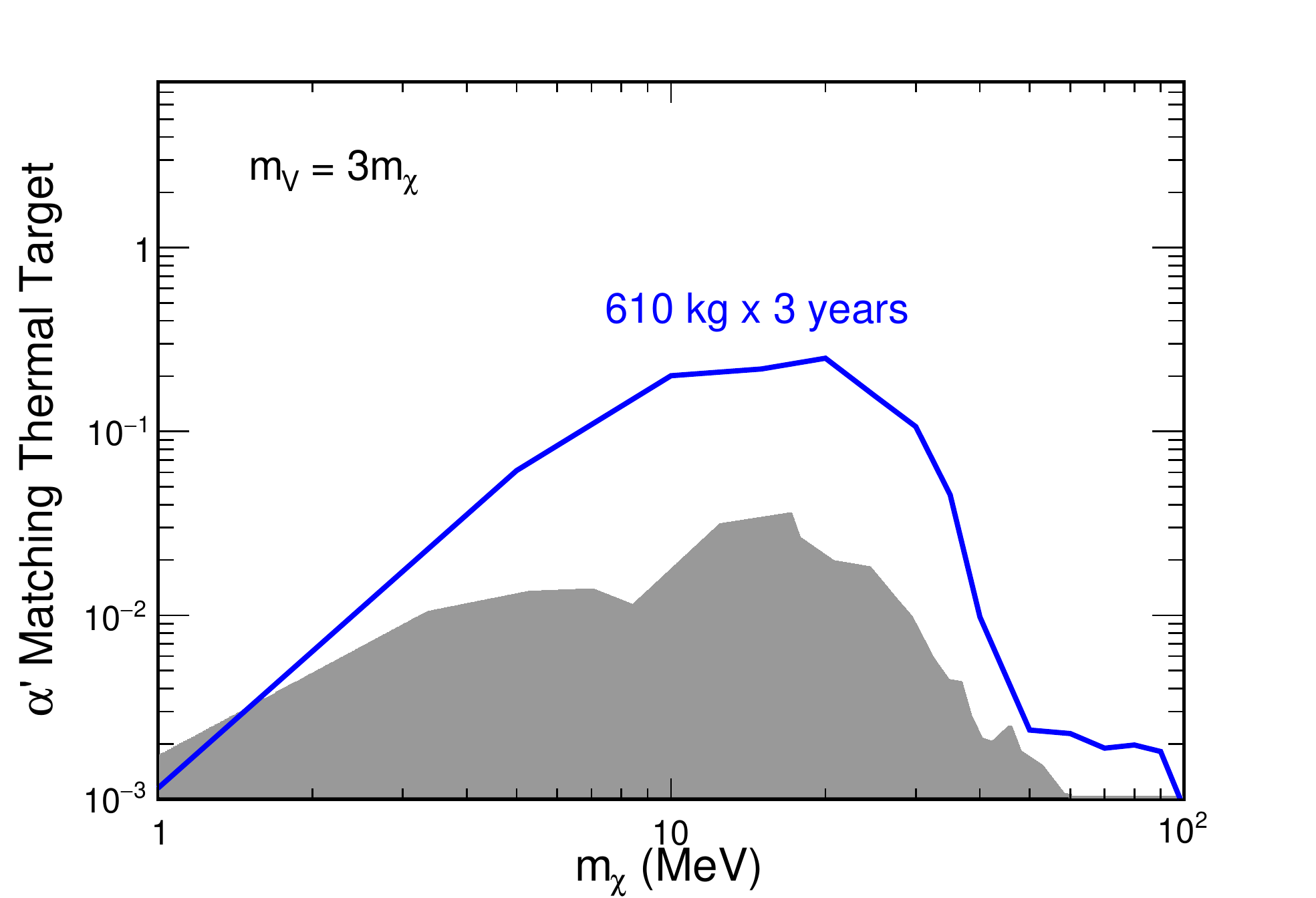}
\caption{Expected sensitivity to constraining dark matter with a vector portal with three years of LAr-1t exposure, with currently excluded regions in grey.  The top plot gives the limits on $Y$ compared to the thermal flux of dark matter, assuming $\alpha'=0.5$.  The bottom plot shows the values of $\alpha'$ for which the thermal flux would be excluded.}
\label{fig:VanillaVectorSensitivities}
\end{figure}

For each point in parameter space, the $\Delta\chi^2$ between the tested parameters and the standard model without dark matter is calculated.  The 90$\%$ sensitivity curve, drawn using $\Delta\chi^2<2.706$, constraining a vector portal to sub-GeV dark matter is shown in Fig \ref{fig:VanillaVectorSensitivities}.  Assuming the model mediated by a vector portal particle that kinematically mixes with the photon, we improve constraints for $1.3$~$<$~$m_\chi$~$<$~$70$~MeV$/c^2$, nearly the entire range accessible to the detector.  For a choice of mass parameters, the observed flux of cosmological dark matter scales is expected at a specific $Y$, while the event rate in our detector scales like $\epsilon^4\alpha'\propto Y^2/\alpha'$ \cite{Izaguirre:2015yja}.  Thus, any constraint on $Y$ can be translated to an upper constraint on $\alpha'$, below which the experiment would rule out the cosmologically observed dark matter flux within the model.  Also, the sensitivity to exclude a certain flux of dark matter is related linearly to the sensitivity in $\alpha'$.  For $6 < m_\chi < 30$~MeV$/c^2$, we would rule out the thermal flux for $\alpha'<0.1$.  For much of the studied range, we can improve on current constraints by over an order of magnitude on dark matter flux.  Currently, this kinematically-mixed vector model is best constrained with data from LSND for $m_\chi<8$~MeV$/c^2$~\cite{Auerbach:2001wg}, MiniBooNE for $8<m_\chi<60$~MeV$/c^2$~\cite{Aguilar-Arevalo:2017mqx}, and BaBar \cite{Aubert:2008as} for $m_\chi>60$~MeV$/c^2$ in the parameter space of interest with data from E137~\cite{PhysRevD.38.3375} and BNL~E949~\cite{Artamonov:2009sz} competitive.

\begin{figure}[!bt]
\centering
\includegraphics[width=0.47\textwidth]{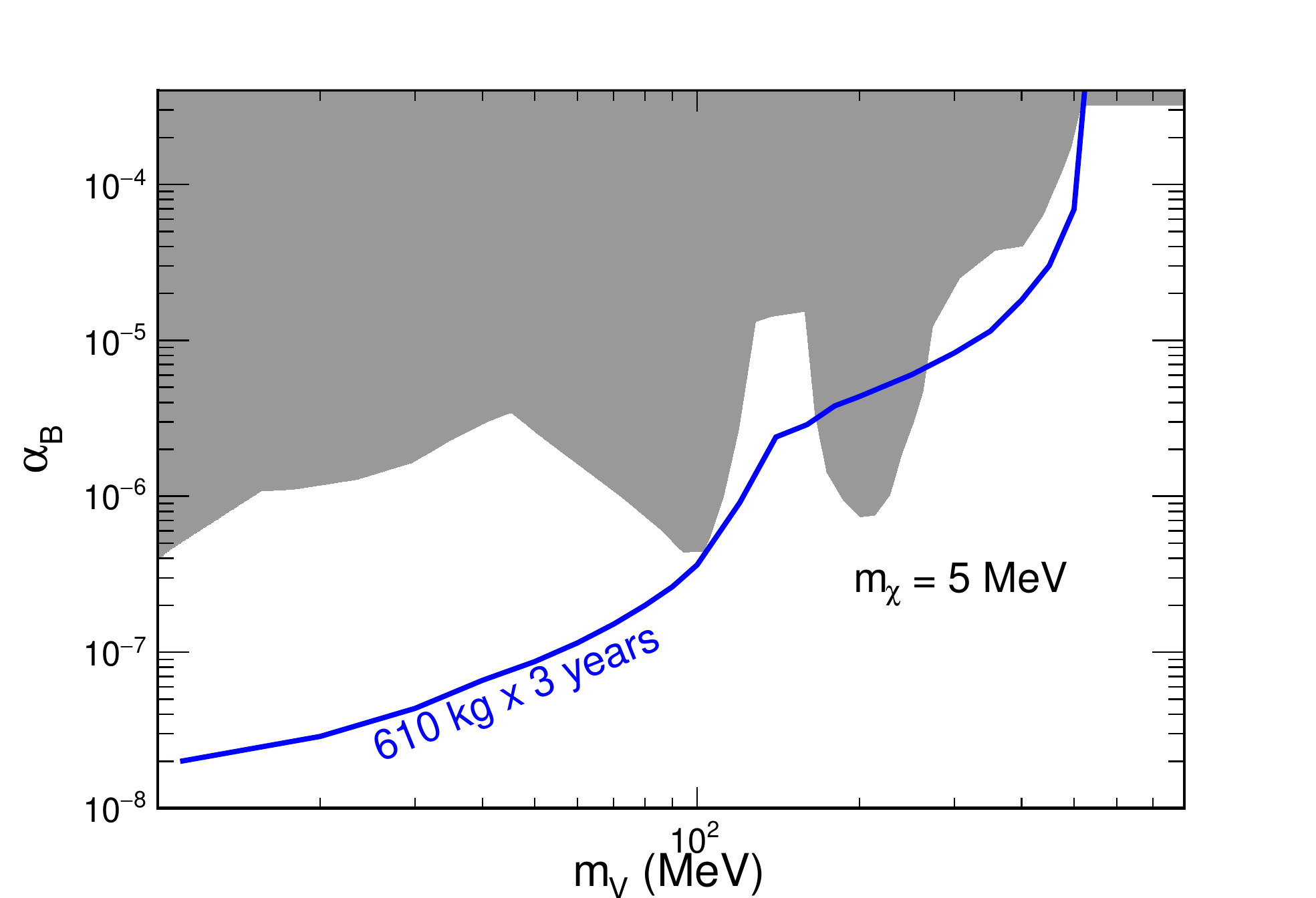}
\caption{Expected sensitivity to constraining dark matter with a leptophobic portal with three years of LAr-1t exposure, with currently excluded regions in grey.  All constraints are shown as a function of $m_V$ with $m_\chi$~=~$5$~MeV$/c^2$.}
\label{fig:VanillaBarySensitivities}
\end{figure}

As the current best constraints on this dark matter model mediated by a vector are incapable of constraining the leptophobic portal model, our sensitivity to this model is very competitive in the region between 10 and 500 MeV, as shown in Fig. \ref{fig:VanillaBarySensitivities}.  This improves on current bounds of dark matter flux by up to six orders of magnitude as $\phi_{DM}\propto\alpha_B^3$.

\section{Future Prospects for \cevns\ Detectors}
\label{sec:Future}
\begin{figure}[!bt]
\centering
\includegraphics[width=0.49\textwidth]{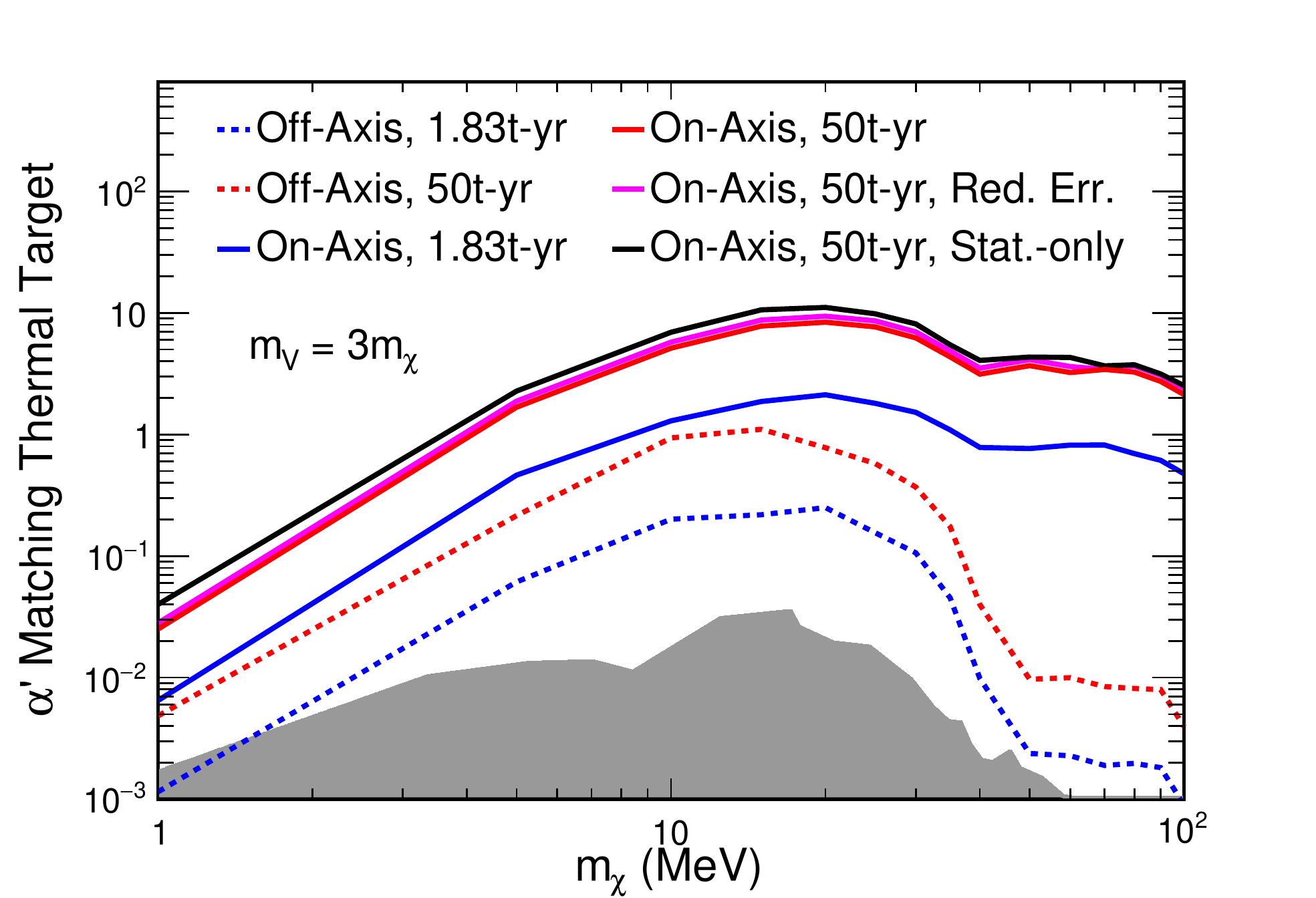}
\caption{The reach of a dark matter search with a next-generation liquid argon \cevns\ detector with various potential improvements as described in the text.  Implementing a 10-tonne scale would push constraints, but an on-axis detector is needed to rule out all perturbative values of $\alpha'$.  The perturbative range is ruled out for a 10-tonne on-axis detector for $4<m_\chi<100$~MeV$/c^2$ in the analyzed window.}
\label{fig:ReachSensitivities}
\end{figure}

\cevns\ detectors can significantly improve on the LAr-1t sensitivity with realistic assumptions on detector configuration and systematic errors that are achievable with the next generation of detectors.  We estimate that a liquid argon detector could probe a dark matter flux up to 2000$\times$ lower than current constraints.  A comparison of potential improvements is shown in Fig. \ref{fig:ReachSensitivities}, showing \cevns\ detectors have the potential to cover the perturbative region of parameter space with $\alpha'<1$ for $4 < m_\chi < 100$~MeV$/c^2$ within the vector portal model.

A detector capable of determining the directionality of any observed nuclear recoil signal \cite{Kadribasic_2018} from dark matter may further improve background rejection techniques.  Additionally, a confirmation of the angular differential cross section would serve as a valuable check for confirming any observed excess is consistent with dark matter scatters.  A detector with such capabilities would be very different from the proposed scintillation detector described here, though its sensitivity in the SNS beam line would be an interesting future calculation.

\subsection{Effectiveness of Analysis Strategy with Higher Mass Detectors}

As shown in Fig. \ref{fig:DMErrorBand}, statistical errors dominate at 610 kg $\times$ 3 years of exposure.  We therefore would expect continued improvement in dark matter searches with a larger accumulated dataset.  With a next-generation detector with 10 tonnes of fiducial volume, sensitivity would continue to improve after several years of exposure.  After 50 tonne-yr of exposure, this measurement would not yet be systematically limited.

\subsection{Reduction of Flux Uncertainty}

In the years before LAr-1t is commissioned, understanding of the relevant systematic uncertainties is likely to improve.  Thus, we repeat the above sensitivity calculation with assumptions of reduced errors.  The neutrino and dark matter flux uncertainty is reduced from 10$\%$ to $3\%$, which we plan to achieve through an independent measurement of the neutrino flux using a D$_2$O detector in Neutrino Alley.  Using the precisely calculated $\nu_e$ CC cross section on deuterium~\cite{PhysRevC.63.034617}, the measured event rate would give the neutrino flux with small uncertainty.  The enhancement in sensitivity we achieve with this reduced uncertainty is shown in Fig. \ref{fig:ReachSensitivities}.  The sensitivity improvement is equivalent to a roughly 30$\%$ increase in accumulated exposure for each of the scenarios tested.

\subsection{Utilizing Angular Dependence of the Dark Matter Flux}

\begin{figure}[!bt]
\centering
\includegraphics[width=0.49\textwidth]{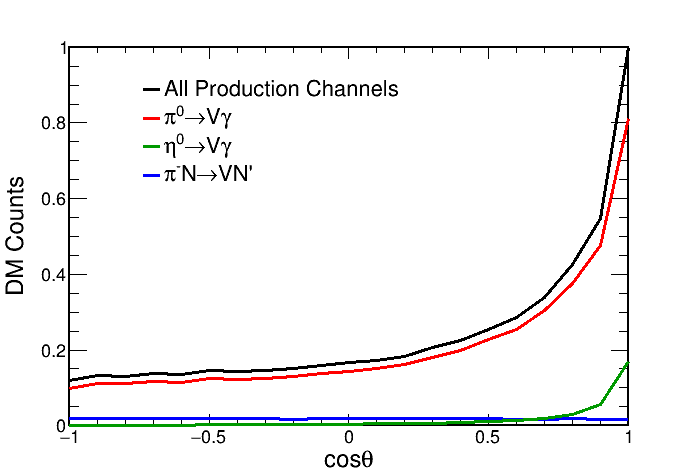}
\caption{The expected rate of dark matter scatters in our detector as a function of the placement angle relative to the beam, as a fraction of the on-axis flux and separated by production channel with $m_\chi = 15$ MeV and $m_V = 45$ MeV.  For portal masses below the pion mass, there are notable contributions from three production channels, but for $m_\chi>m_\pi/2$, $\eta\rightarrow V\gamma$ dominates.}
\label{fig:DMAngularependence}
\end{figure}

Though the neutrino flux from a $\pi$-DAR beam is isotropic, the flux of portal particles to the dark sector produced would be boosted, and thus correlated with the beam direction.  A \cevns\ detector placed on-axis would thus see a greater dark matter flux, increasing the predicted signal.  The angular dependence of the predicted dark matter flux through the detector is shown in Fig. \ref{fig:DMAngularependence} with the baseline fixed to the LAr-1t planned location: 28.4 m.

In addition to increased flux produced through $\pi^0\rightarrow V\gamma$, there is a notable contribution from $\eta^0$~$\rightarrow$~$V\gamma$.  These are produced near threshold and are non-relativistic in the center-of-momentum frame and thus boosted into very forward directions in the lab frame.  In the parameter space $m_\pi/2$~$<$~$m_\chi$~$<$~$m_\eta/2$, where there is no production through $\pi^0$ decay, the increase in flux can be quite large.  For instance, we expect a 5.5$\times$(390$\times$) increase in dark matter flux on-axis with $m_\chi=$~1(100)~MeV$/c^2$ and $m_V=3(300)$~MeV$/c^2$.  Also, in the event of an experimental detection of sub-GeV dark matter, mapping this angular dependence with a \cevns\ detector would be key evidence to strengthen the claim of discovery and refine understanding of physics parameters.

With the trade-off between the increase in neutron flux on-axis and more strategies to reduce the neutron flux in a purpose-built detector hall, we assume the neutron rate in the detector is the same as in the CENNS-10 position in Neutrino Alley.  With this and the LAr-1t steady-state and \cevns\ backgrounds, we assemble a sensitivity for an on-axis detector with both 610 kg $\times$ 3 yrs and 10-t $\times$ 5 yrs of exposure.  In either case, the on-axis detector has the potential to probe $\phi_{DM}$ 6$\times$ lower than that accessible to a detector in the CENNS-10 location, with the gains yet more significant for $m_\chi>m_\pi/2$.

\section{Conclusions}
\label{sec:Conclusion}
Dark matter direct detection experiments have made significant progress searching for a flux of thermal WIMP particles in the past decades.  However, these experiments quickly lose sensitivity for dark matter masses below the Lee-Weinberg bound, $\sim$1 GeV$/c^2$.  Masses below this, however, are well within the grasp of accelerator-based searches.  Thus, direct-detection and detection of accelerator-produced dark matter probe complementary mass scales and together can extend the probed parameter space for a relic dark matter WIMP.  

We assess the ability for the LAr-1t detector, whose design is being finalized and is to be implemented at the SNS, to probe vector and leptophobic portals to sub-GeV scalar dark matter.  With uncertainties informed by our ongoing experience with LAr \cevns\ detectors, we find such a detector would improve upon current bounds for $1$~$<$~$ m_\chi$~$<$~$100$~MeV$/c^2$.  The coherent enhancement of the \cevns-like dark matter cross section on nuclei and ability to constrain systematic uncertainties using the delayed \cevns{} from a $\pi$-DAR beam gives us leading sensitivity even compared to detectors more than an order of magnitude larger.

The LAr scintillation detector technology, already being studied by the collaboration with a small prototype detector, is an ideal vehicle to test this dark matter parameter space.  This detection strategy is easily scalable, with a 50 tonne-yr exposure reasonably achievable with a next-generation detector.  Through implementing a 10 tonne-scale detector while taking advantage of the directionality of the dark matter flux, these detectors would test parameter space that are consistent with the observed flux of astrophysical dark matter.  The sensitivity reach of this strategy is significant, for a large range of dark matter masses would represent a test of dark matter flux three orders of magnitude below the current best limits.

\section{Acknowledgments}

The COHERENT collaboration acknowledges the generous resources provided by the ORNL Spallation Neutron Source, a DOE Office of Science User Facility, and thanks Fermilab for the continuing loan of the CENNS-10 detector. We also acknowledge support from the Alfred~P. Sloan Foundation, the Consortium for Nonproliferation Enabling Capabilities, the Institute for Basic Science (Korea, grant no. IBS-R017-G1-2019-a00), the National Science Foundation,  the Russian Foundation for Basic Research (proj.\# 17-02-01077 A), and the U.S. Department of Energy, Office of Science. Laboratory Directed Research and Development funds from ORNL and Lawrence Livermore National Laboratory also supported this project.  This research used the Oak Ridge Leadership Computing Facility, which is a DOE Office of Science User Facility.  We thank Bhaskar Dutta, Louis Strigari, Doojin Kim, and Shu Liao for useful discussions.

\bibliographystyle{apsrev4-1}
\bibliography{main}

\end{document}